\documentclass[aps,prl,twocolumn,showpacs,floatfix]{revtex4}
\usepackage{graphicx}
\usepackage{psfrag}
\usepackage{color}

\newcommand{\ie}{\emph{i.}$\,$\emph{e.}}

\def\vec #1{{\bf #1}}

\begin{document}
\title{Self-Contact and Instabilities in the Anisotropic Growth of Elastic Membranes}

\author{ Norbert Stoop$^1$, Falk K.\ Wittel$^1$, Martine Ben Amar$^2$, Martin Michael M\"uller$^{2,3}$, and Hans J.\ Herrmann$^{1,4}$} 
\affiliation{$^1$\ Computational Physics for Engineering Materials, ETH Zurich, Schafmattstr. 6, HIF, CH-8093 Zurich, Switzerland\\
$^2$\ Laboratoire de Physique Statistique de l'Ecole Normale Sup\'erieure (UMR 8550), associ\'e aux Universit\'es Paris 6 et Paris 7 et au CNRS; 24, rue Lhomond, 75005 Paris, France\\
$^3$\ Laboratoire de Physique Mol\'eculaire et des Collisions, Equipe BioPhysStat, ICPMB-FR CNRS 2843, Universit\'e Paul Verlaine-Metz; 1, boulevard Arago, 57070 Metz, France\\
$^4$ Departamento de F\'isica, Universidade Federal do Cear\`a, Campus do Pici, 60451-970 Fortaleza, Cear\'a, Brazil}
\date{\today}

\begin{abstract}
We investigate the morphology of thin discs and rings growing in the circumferential direction. Recent analytical results suggest that this growth produces symmetric \textit{excess} cones (\textit{e}-cones). We study the stability of such solutions considering self-contact and bending stress. We show that, contrary to what was assumed in previous analytical solutions, beyond a critical growth factor, no symmetric \textit{e}-cone solution is energetically minimal any more. Instead, we obtain skewed \textit{e}-cone solutions having lower energy, characterized by a skewness angle and repetitive spiral winding with increasing growth. These results are generalized to discs with varying thickness and rings with holes of different radii.
\end{abstract}

% 05.45.-a Nonlinear dynamics and chaos
% 68.35.Rh Phase transitions and critical phenomena
% 89.75.Da Systems obeying scaling laws

\maketitle
%%%%%%%%%%%%%%%%%%%%%%%%%%%%%%%%%%%%%
%\section{Introduction}
Growing membranes are ubiquitous in living matter, producing a very rich variety of complex shapes. In an attempt to sustain external loads and to minimize internal stresses, structures adopt new shapes. Non-uniform or restrained growth of surfaces inevitably leads to out of plane buckling. This can be evidenced in many examples such as the development of the folds of \textit{Mitochondria}, the formation of plant leaves, the wrinkling of human skin or, in two dimensions, the packing of rod-like objects \cite{nstoop}. 

Flat discs growing under various conditions have received intense attention recently from experimental \cite{sharon1,sharon2} and theoretical sides \cite{santangelo, julien, mmm}. Santangelo \cite{santangelo} presented a theoretical study to explain experiments on radially symmetric, isotropic growth of polymers with different swelling factors \cite{sharon1,sharon2}. In Dervaux et al. \cite{julien}, potato-chip like shapes were found for anisotropic, circumferential growth using modified F\"oppl-von K\`arm\`an equations \cite{julien2}. This description is however only valid for small deformations and was extended in Ref.~\cite{mmm} to large deformations in the limit of vanishing thickness. It was found that circumferential growth is equivalent to inserting a wedge into the disk, leading to so-called \textit{excess} cones or "\textit{e}-cones". They are characterized by an excess angle $\varphi_e$ which is the angle of the wedge added to the disc. With increasing $\varphi_e$, the symmetric \textit{e}-cone will eventually touch itself simultaneously with two perpendicular contact planes. From that situation on, the authors of Ref.~\cite{mmm} made the assumption that the \textit{e}-cone remains symmetric, and derived the corresponding bending energy.

In this Letter, we show that this assumption is not correct when $\varphi_e$ is above a critical value. Instead, by taking surface self-contact into account, a family of new shapes emerges, namely skewed \textit{e}-cone solutions with a varying contact plane angle. Our results are obtained from a sophisticated numerical thin shell model capable of capturing large deformation and growth.  We find the energetically most favorable shapes, discuss the dependence of stable morphologies on shell thickness, and highlight the analogy between disk and ring growth.

%%%%%%%%%%%%%%%%%%%%%%%%%%%%%%%%%%%%%
%\section{Mechanics of thin shells}
The mechanics of thin discs is well described by the classical Kirchhoff-Love model. It follows by expansion of the full 3d elastic problem in the thickness $h$, and by assuming negligible transverse shear through the thickness \cite{landau_lifshitz, benamarpomeau, friesecke}. The elastic energy is then entirely given by the deformation of the disc's middle-surface $\Omega$, and contains a stretching and a bending part:
\begin{eqnarray}\label{energy}
E &=& \int_\Omega \left[\frac{Y h}{2 (1-\nu^2)} U_s 
  + \frac{Y h^3}{24 (1-\nu^2)} U_b \right] \, d\Omega \; .
\end{eqnarray}
Here, $Y$ is the elastic modulus, $\nu$ Poisson's ratio and $d\Omega$ the infinitesimal area element. The stretching and bending energy densities $U_s$ and $U_b$ are given by
\begin{eqnarray}\label{energydensity}
U_s &=& (\epsilon_{11}+\epsilon_{22})^2 - 2(1-\nu)(\epsilon_{11}\epsilon_{22}-\epsilon_{12}^2) \quad \text{and} \quad\;\;\;\;\;\\
U_b &=& (\kappa_{11}+\kappa_{22})^2 - 2(1-\nu)(\kappa_{11}\kappa_{22}-\kappa_{12}^2) \; .\label{bendingdensity}
\end{eqnarray}
$\epsilon_{ij}$ denote the in-plane strains and $\kappa_{ij}$ the components of the curvature tensor. The first term in $U_b$ is the squared mean curvature, while the second term accounts for Gaussian curvature.
%%%
%%%%%%%%%%%
\begin{figure}[ht!!!!]
%%\epsfxsize=5.5cm
%%\centerline{\epsffile{snapshotsdisk.eps}}
%\epsfxsize=8.6cm
%\centerline{\epsffile{/Users/norbert/Desktop/Econe_Figures/econes.eps}}
%\centerline{\epsffile{fig1.eps}}
\centerline{\includegraphics[width=8.6cm]{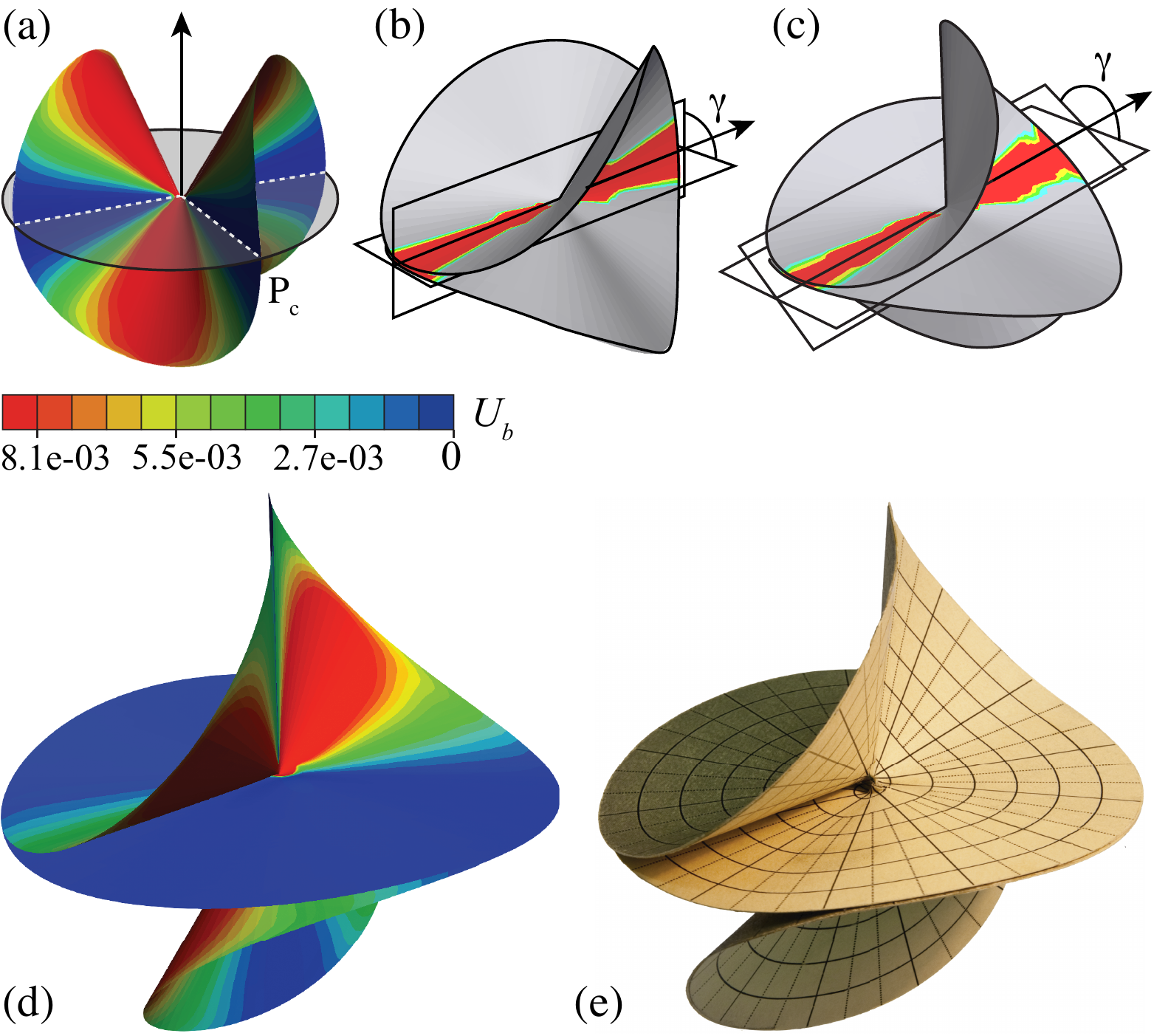}}
\caption{\label{fig1} Numerical equilibrium solution for the touching problem of \textit{e}-cones: Increasing the excess angle $\varphi_e$ of a symmetric \textit{e}-cone (a), the four wings start to touch, initially with angle $\gamma=\pi/2$ between the two touching planes (b). For larger $\varphi_e$, the angle changes (c) and a skewed \text{e}-cone emerges (d), where the contacting wings have slided against each other. A paper model is shown in (e) for comparison. Colors in (a) and (d) represent the bending energy density $U_b$. Red color in (b) and (c) denotes points with self-contact. Simulation parameters are $Y=10^4\,$Pa, $\nu=0.3$, $R=1\,$m, $h=0.02\,$m (lenghts are rescaled by $R$ in the main text). Values of $\varphi_e$ for (a)-(e) are $(3.73, 7.45, 7.6, 4\pi, 4\pi)$. Dashed lines in (a) denote points $P_{\text{c}}$ of vanishing curvature. For $\varphi_e=3.73$, their tangents are perpendicular to the gray-shaded plane.} %Supplementary video available online.}
\end{figure}
%%%%%%%%%%%
%%%
Without growth, $\epsilon = 1/2 (F^TF - 1)$, where $F=\nabla\chi$ is the deformation gradient of the map $\chi$ that transforms from the reference to the deformed configuration, see e.g. Ref.~\cite{ogden}. Growth is added by decomposing $F$ multiplicatively into $F = AG$ \cite{rodrigues,goriely}. $G$ is the growth tensor describing the change of mass and $A$ is the elastic tensor that ensures compatibility (no overlap) and continuity (no cavitation) of the body. We further assume: \textit{(i)} Growth is slow compared to the elastic timescale, which is typically true in nature \cite{goriely}. Thus, the disc is always in elastic equilibrium. \textit{(ii)} There exists a stress-free reference state, \ie\ the flat disc with $G=\mathbf{1}$. \textit{(iii)} The elastic response depends only on $A$. The in-plane strains then become $\epsilon = 1/2(A^TA-1)$ and $\kappa$ is modified similarly. In the present model, we only consider surface expansion, \ie, the thickness remains constant. In terms of the excess angle $\varphi_e$, $G$ in polar coordinates then becomes
\begin{equation}
G(r,\varphi) = 	\left( \begin{array}{cc}
						1 & 0 \\
						0 & 1 + \varphi_e/(2 \pi)
				\end{array} \right) \; .
\end{equation}
%%%%%%%%%%%%%%%%%%%%%%%%%%%%%%%%%%%%%
%\section{Simulating growth of shells}
We search for configurations that minimize Eq.~(\ref{energy}) by discretizing the system using the finite element method. The curvature integral in Eq.~(\ref{energy}) requires at least continuity of first derivatives. To this end, we employ a recently developed method based on $C^1$-continuous subdivision finite elements (SDFEs) \cite{cirak}. From the discrete Lagrangian formulation of Eq.~(\ref{energy}) the elastic forces $\vec{f}_i$ are derived for all mesh nodes $i$, and the equations of motion are integrated in time. For numerical stability and equilibration, we add an external viscous damping term of the form $\eta\vec{v}_i$ to the equations of motion, with $\vec{v}_i$ the velocity of node $i$. $\eta$ was chosen subcritical for all vibration modes and did not have influence on the results.
To validate our model, we use the non-linear ABAQUS software package with implicit solver and 4 node general shell elements (S4) including transverse shear \cite{abaqus}. Contact is modeled in ABAQUS via surface-surface contact and a linear force law while in our implementation, point-triangle contacts are subject to a quadratic repulsion force. Note that in tangential direction contacts are frictionless unless explicitly mentioned. 

The simulation starts from a flat disc with constant thickness $h$, with $h$ between $0.001$ and $0.04$. The excess angle $\varphi_e$ is increased in small steps by expanding all elements in the tangential direction of a centered cylindrical material coordinate system. Assuming slow growth, the equilibrium solution is obtained after every such step (in SDFEs, we wait until the kinetic energy becomes negligible). In SDFE simulations a small vertical random noise is initially imposed on the displacements to break the planar symmetry of the flat disk. In ABAQUS, this is automatically provided by numerical noise. In dynamic simulations via ABAQUS, we first observe high buckling modes of the \textit{e}-cone type, and for $h$ small enough even superimposed radial buckles. Due to the bending rigidity these soon damp out into the lowest energy mode with two folds ($n$=2), having a shape similar to a potato chip as shown in Fig.~\ref{fig1} (a). %The relevant time scale to reach the $n$=2 mode is given by the shell thickness.
The growth continues stable in the $n$=2 mode. The analytical solution predicts contact between the two folds at $\varphi_{e}^{\text{kiss}} = 7.08$ \cite{mmm}. We find excellent numerical agreement, with the first contact occurring between $\varphi_{e} = 7.12\pm0.01$ for $h=0.04$, and $\varphi_{e}=7.074\pm0.005$ for $h=0.001$.

The analytical solution suggests an increase in bending energy up to the theoretical value of the solution with three wings ($n$=3), followed by a transition into this mode \cite{mmm}. We obtain a flattening of the contact zones maintaining symmetry in two planes (Fig.~\ref{fig1} (b)). Shortly after reaching $\varphi_e^{\text{kiss}}$, however, the contact planes start to counter rotate (Fig.~\ref{fig1} (c)), leading to a reduction of curvature of one wing at the expense of the other. Consequently a skewed \textit{e}-cone solution with a flat disc-like part and two wings forms as shown in Fig.~\ref{fig1} (d). This transition is obtained for all shell thicknesses considered. If growth continues, we simply keep on adding additional windings to both sides of the flattened region. In the following we address the nature of the transition, the composition of elastic energies, and their dependence on system parameters. 

%%%%%%%%%%%%%%%%%%%%%%%%%%%%%%%%%%%%%
%\subsection{Energy of the finite-thickness \textit{e}-cone}
\par
The analytical results in Ref.~\cite{mmm} are based on discs with vanishing thickness. As a consequence, the \textit{e}-cone shape has zero Gaussian curvature everywhere, except at the point-like apex. 
%In particular, geodesics from the apex to the rim on the \textit{e}-cone surface are straight lines. 
By introducing a finite thickness, it is intuitively clear that the singularity at the apex will be flattened in order to reduce bending energy and because the maximal curvature is bounded by the thickness. Consequently, the Gaussian curvature cannot vanish close to the apex leading to local stretching. We call this region the \textit{core} of the \textit{e}-cone and denote its radius by $R_c$. We numerically calculated the \textit{e}-cone equilibrium solutions for different thicknesses and excess angles and found that $R_c \propto h$, a behavior similar to the core size scaling of \textit{developable} cones reported in \cite{benamarpomeau,witten,cerda}.
%%% 
%%%%%%%%%%%
\begin{figure}[t]
%\epsfxsize=8.7cm
%\centerline{\epsffile{/Users/norbert/Desktop/Econe_Figures/ebend_combined_rescaled2.eps}}
%\centerline{\epsffile{fig2.eps}}
\centerline{\includegraphics[width=8.7cm]{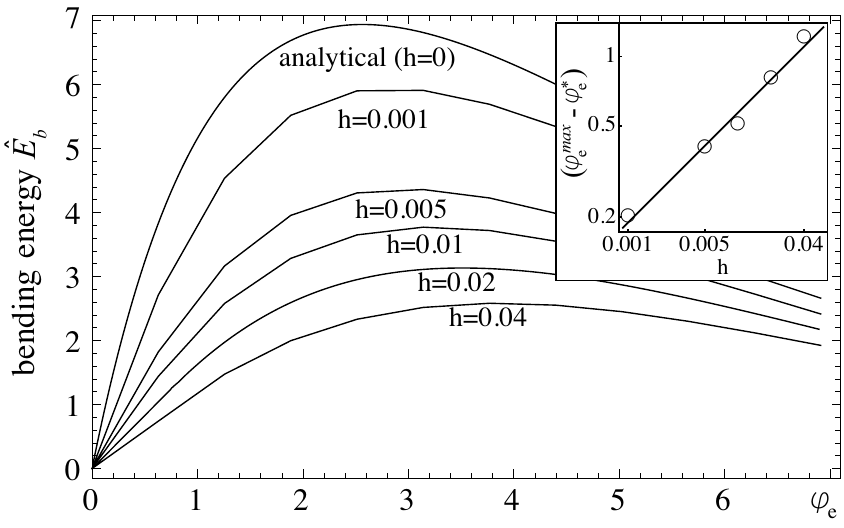}}
\caption{\label{fig3} Normalized bending energy $\hat{E}_b$ in dependence on the disc thickness $h$ for the symmetric \textit{e}-cone, compared to the analytical solution with $h=0$ \cite{mmm}. Inset: The position of the maxima of $\hat{E}_b$, denoted $\varphi_{e}^{\text{max}}$, approaches the analytical value $\varphi_e^*$ according to $(\varphi_e^{\text{max}}-\varphi_e^{*}) \propto h^\alpha$ with $\alpha=0.45\pm0.03$.
}
\end{figure}
%%%%%%%%%%%
%%%
In Fig.~\ref{fig3} we show the rescaled bending energy $\hat{E}_b = \int_\Omega U_b/A \; d\Omega$, where $A$ is the total area of the \textit{e}-cone \footnote{Since stretching occurs, $A$ is not exactly $R^2(2\pi+\varphi_e)/2$.}. The figure shows that the bending energy is indeed strongly reduced for thick shells. To quantify the convergence towards the analytical solution, we consider the excess angle $\varphi_{e}^{\text{max}}$ where $\hat{E}_b$ reaches its maximum. We find that $\varphi_{e}^{\text{max}}$ converges to the analytical maximum $\varphi_e^*=2.57$ according to $(\varphi_e^{\text{max}}-\varphi_e^{*}) \propto h^\alpha$ with $\alpha=0.45\pm0.03$ (Fig.~\ref{fig3}, inset).

%%%%%%%%%%%
\begin{figure}[t]
%\epsfxsize=8.7cm
%%\centerline{\epsffile{skewed_energies_rescaled_combined.eps}}
%\centerline{\epsffile{/Users/norbert/Desktop/Econe_Figures/skewed_energies_rescaled_combined.eps}}
%\centerline{\epsffile{fig3.eps}}
\centerline{\includegraphics[width=8.7cm]{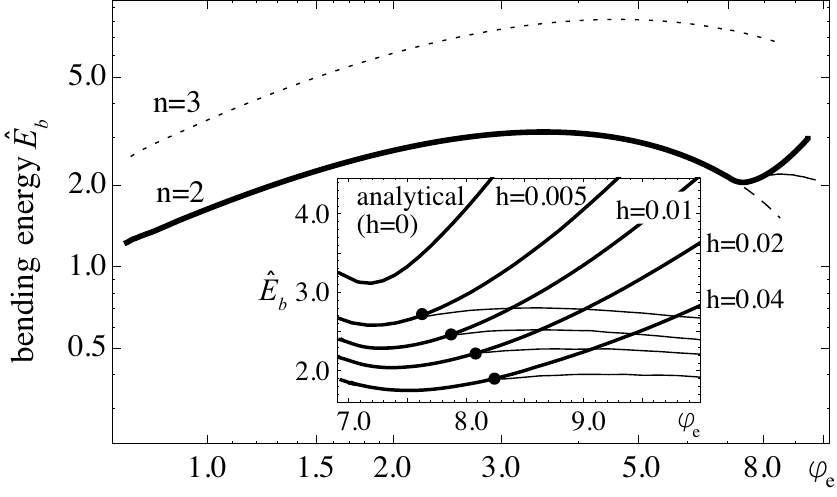}}
\caption{\label{fig4} Bending energies for the $n=2$ contact problem with $h=0.02$: The symmetric $n=2$ \textit{e}-cone (bold line) is stable until $\varphi_e = 8$, when the solution changes into the skewed \textit{e}-cone (thin line). For comparison, the unphysical situation with self-intersections is also shown (dashed line), as well as the $n=3$ solution (dotted line). The inset illustrates the influence of the thickness in the transition region. Bold lines are again the symmetric \textit{e}-cones, and thin lines correspond to the skewed equilibrium solution. The critical excess angles  $\varphi_e^{\text{crit}}$ where \textit{e}-cones loose stability are marked with black dots.}
\end{figure}
%%%%%%%%%%%%

%%%%%%%%%%%%%%%%%%%%%%%%%%%%%%%
%\subsection{Energy of skewed \textit{e}-cone}
For the $n=2$ mode at $\varphi_e^{\text{kiss}}$, the wings touch and eventually form a skewed \textit{e}-cone. In order to analyze the stability of the symmetric solution, we performed two types of simulations: First, we "stabilized" the symmetric solution using static friction between the areas in contact and increased $\varphi_e$ to $10$. This way, we can fix the \textit{e}-cone solution, which would be unstable without friction. At $\varphi_e=10$, the contact friction was removed and the system relaxed towards the skewed \textit{e}-cone. We then followed the skewed solution by decreasing the excess angle from $\varphi_{e}=10$ back to $\varphi_{e}^{\text{kiss}}$.
During this procedure, we measured the bending energy, which is the relevant part of the total energy, shown in Fig.~\ref{fig4}. We note that in the range of $\varphi_e$ investigated, the next higher $n=3$ mode (dotted line) is energetically above the symmetric \textit{e}-cone (bold line), despite the fact that it is a configuration that does not involve any contact (for $n=3$, the wings would touch only for $\varphi_e=13.3$). Fig.~\ref{fig4} furthermore shows that above a critical $\varphi_e^{\text{crit}}$, the skewed solution provides the lowest energy, whereas the symmetric \textit{e}-cone is favored in the regime $\varphi_{e}^{\text{kiss}}<\varphi_e<\varphi_e^{\text{crit}}$. The influence of the finite disc thickness on the energy and stability is illustrated in the inset of Fig.~\ref{fig4}. Remarkably, even though $\varphi_e^{\text{kiss}}$ is practically identical for all thicknesses considered, we find a significant dependence of $\varphi_e^{\text{crit}}$ on $h$ (black dots). 
\par
Finding the minimum of the nonlinear problem posed by Eq.~(\ref{energy}) with self-contact is nontrivial and numerically delicate. We therefore compared our numerical findings to experiments by measuring the angle $\gamma$ between the touching planes (cf. Fig.~\ref{fig1}) for h=0.005. 
%%%%%%%%%%%
\begin{figure}[b]
%\epsfxsize=6cm
%%\centerline{\epsffile{snapshotsring.eps}}
%\epsfxsize=8.6cm
%\centerline{\epsffile{/Users/norbert/Desktop/Econe_Figures/rings.eps}}
%\centerline{\epsffile{fig4.eps}}
\centerline{\includegraphics[width=8.6cm]{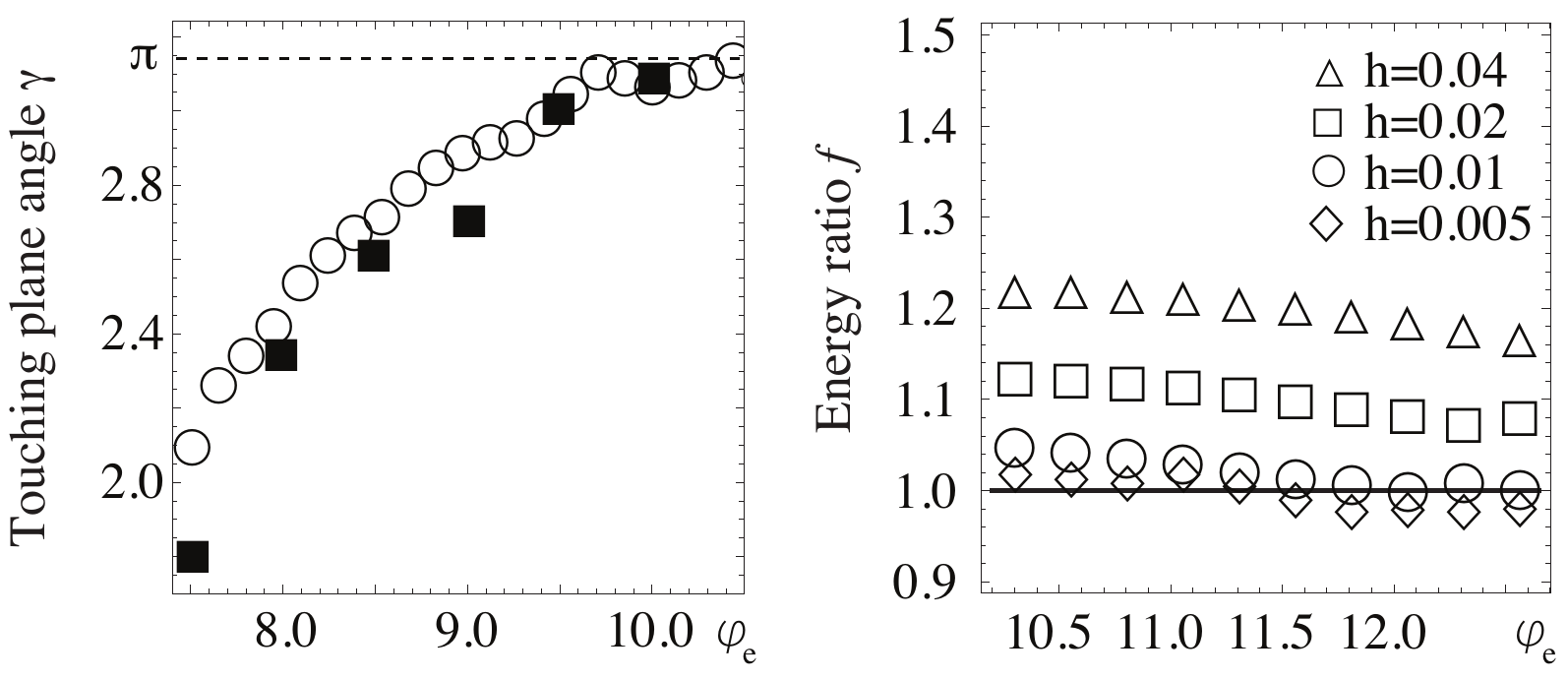}}
\caption{\label{fig5} Left: Dependence of the touching plane angle $\gamma$ (cf. Fig.~\ref{fig1}) on $\varphi_e$, determined numerically (circles) and experimentally (squares) for $h=0.005$. Right: Bending energy ratio $f=\hat{E}_b(\varphi_e^{\text{touch}})/\hat{E}_b(\varphi_e = 3.73)\cdot(2\pi+\varphi_e^{\text{touch}})/(2\pi+3.73)$ between  skewed \textit{e}-cones and the symmetric one at $\varphi_e=3.73$.}
\end{figure}
%%%%%%%%%%%
In simulations, this was done by averaging over the contact forces in each of the two contact zones, resulting in an approximation for the contact plane normals from which the angle $\gamma$ is obtained. Experimentally, we used two layers of paper glued together to obtain a homogenous bending stiffness. The angle was obtained from digital camera images taken from the top view of the \textit{e}-cone. The results in Fig.~\ref{fig5} (left) show good agreement of simulations and experiments, confirming the validity of the model (\ref{energy}) and the numerical equilibrium solution.

Moreover, for values of $\varphi_e>10$ the skewed e-cone solution approaches a shape which can be approximated by a flat disk with two connecting loops (see Fig.~\ref{fig1} d). Both loops put together are described remarkably well by the symmetric free \textit{e}-cone shown in Fig.~\ref{fig1} a: 
From the shape equations of the symmetric \textit{e}-cone it can be derived that for $\varphi_e = 3.73$, the tangent in the points $P_{\text{c}}$ of vanishing curvature is perpendicular to the plane in which the $P_{\text{c}}$ lie. These points are in fact equivalent to the points of contact of the skewed e-cone. The described approximation yields an estimate for the bending energy. 
One obtains for a skewed e-cone with excess angle $\varphi_e^{\text{touch}}$: $\hat{E}_b(\varphi_e^{\text{touch}})\approx\hat{E}_b(\varphi_e = 3.73)\cdot (2\pi+3.73)/(2\pi+\varphi_e^{\text{touch}})$. 
In Fig.~\ref{fig5} (right), we plot the ratio of the two energies $\hat{E}_b(\varphi_e^{\text{touch}})/\hat{E}_b(\varphi_e = 3.73)\cdot(2\pi+\varphi_e^{\text{touch}})/(2\pi+3.73)$, which is indeed close to the predicted value (solid line). Understandably, the quality of the approximation decreases with larger $h$, as the change in bending energy of the \textit{e}-cone core region is not taken into account.

\begin{figure}[t]
%\epsfxsize=6cm
%%\centerline{\epsffile{snapshotsring.eps}}
%\epsfxsize=8.6cm
%\centerline{\epsffile{/Users/norbert/Desktop/Econe_Figures/rings.eps}}
%\centerline{\epsffile{fig5.eps}}
\centerline{\includegraphics[width=8.6cm]{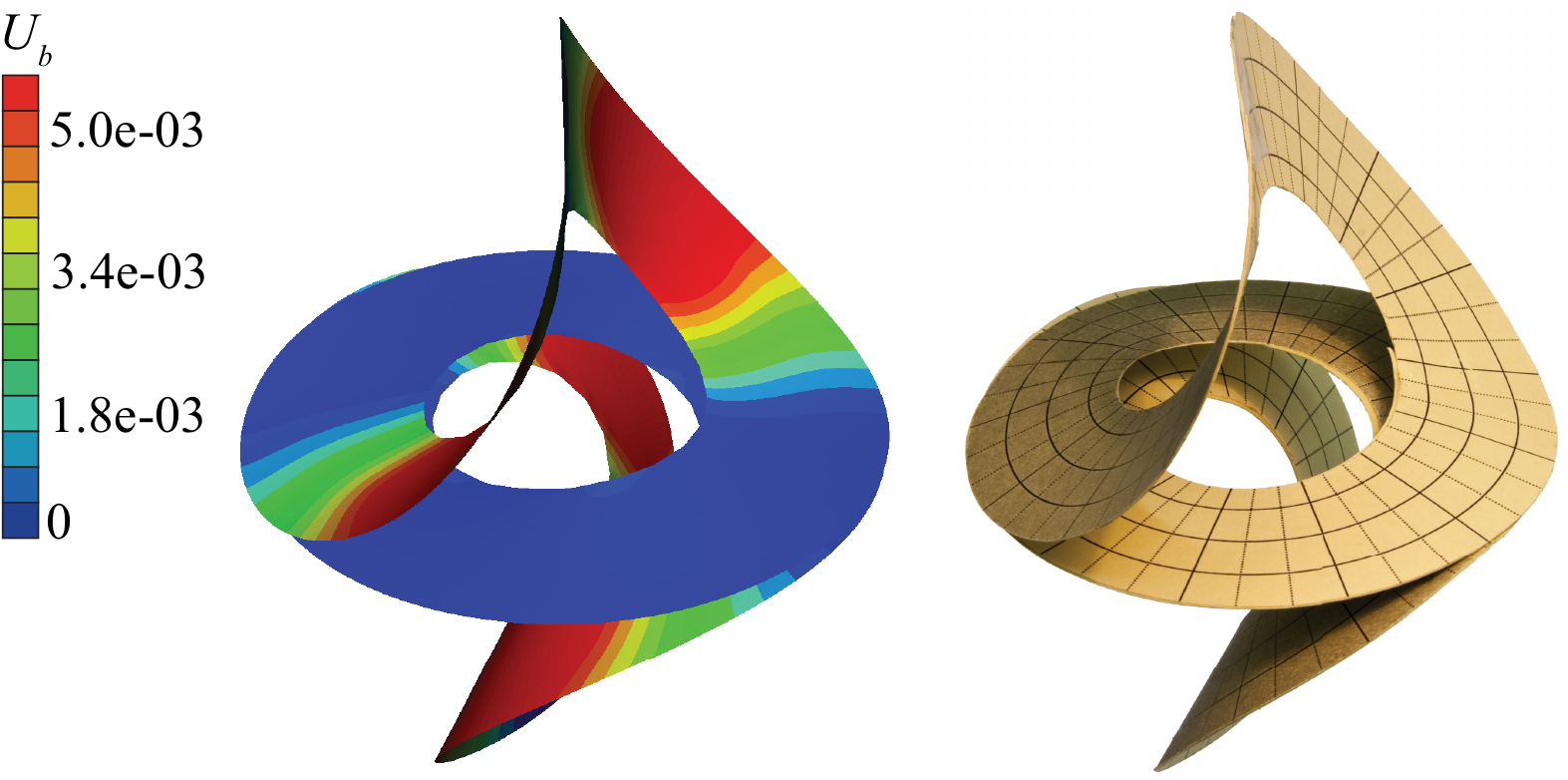}}
\caption{\label{fig6} Numerical equilibrium solution (left) for a growing ring compared to a paper model (right) at $\varphi_e=4\pi$. Colors represent the bending energy density $U_b$. Simulation parameters are identical to Fig.~\ref{fig1}, with $R_i=0.5$.}
\end{figure}

%\subsection{Behavior of rings}
We finally turn to the study of rings with inner radius $R_i$. The same simulation procedure was applied with fixed thickness $h=0.02$. We found that rings, like discs, loose stability after contact occurs and attain the skewed solution  shown in Fig.~\ref{fig6}. Interestingly, $\varphi_e^{\text{kiss}}$ strongly depends on $R_i$, as shown in Table~\ref{table1}. In terms of energy and stability, rings behave similar to discs.

%%%%%%%%%%%
%%%%%%%%%%%

%%%%%%%%%%%
\begin{table}[t!!!!]
\begin{tabular}{c|ccc}
	\hline\hline
	 &  $R_i=0.1$    &   $R_i=0.25$  & $R_i=0.5$   \\
	\hline
$\varphi_e^{\text{kiss}}$ & $7.3\pm0.05$ & $7.8\pm0.05$ & $8.45\pm0.05$ \\
\hline\hline
\end{tabular}	
\caption{\label{table1} Dependence of $\varphi_e^{\text{kiss}}$ on the inner radius $R_i$ of a ring for $h=0.02$.}
\end{table}
%%%%%%%%%%%
%%%%%%%%%%%%%%%%%%%%%%%%

%\section{Conclusions}
We presented a realistic simulation of circumferential growth of discs and rings with finite thickness. By including self-contact, asymmetric, skewed morphologies, that were not predicted by previous analytical solutions, emerge and stabilize for sufficient growth. By energetical arguments based on symmetric deformation modes, we showed, that the asymmetric mode has lower bending energy, and is therefore preferred. Moreover, we found that tangentially growing rings of various widths behave analogous to discs. Good agreement with experiments suggests that the skewed solutions are energetically optimal, but a rigorous proof remains an open question. The asymmetries arising from circumferential growth disclosed in the present work constitutes a novel step in understanding the shapes of membranes and sheets. One can generalize our work to non-linear material behavior and stress or strain dependent growth laws in order to tune to more specific applications in Nature. 
%%%%%%%%%%%%%%%%%%%%%%%%

%\subsection{Acknowledgement} 
This work was supported by Grant No. TH-06 07-3 of the Swiss Federal Institute of Technology Zurich and FUNCAP. The authors would like to thank J. Guven for helpful discussions.

%%%%%%%%%%%%%%%%%%%%%%%%

%%%%%%%%%%%%%%%%%%%%%%%%%%%%%%%%%%%

\end{document}